\documentstyle[preprint,aps]{revtex}

\begin{document}
\draft
\title{Critical phenomena in superlattices: Reentrant dimensional crossover
and anomalous critical amplitudes}
\author{Lev V. Mikheev}
\address{Nordita, Blegdamsvej 17, DK-2100, Copenhagen \O, Denmark}
\date{January 1995}
\maketitle
\begin{abstract}
A crossover from $d$ to $d-1$, and then back to $d$-dimensional critical
behavior is argued to be a generic feature characterizing ordering in a
$d$-dimensional
superlattice composed of atomically {\em thick} films of two ferromagnets.
The crossover
leads to anomalous changes in the amplitudes of critical singularities.
In $d=3$ Heisenberg
and $XY$ superlattices large scale critical fluctuations persist over a wide
temperature
range.
\end{abstract}

\section{Introduction and the statement of the problem}
Recent progress in creating artificial magnetic heterostructures \cite{jbohr}
gives an impetus to the theory of phase transitions in multilayers. While the
related experimental and theoretical efforts have been so far mostly
concentrated
at microscopic effects, this paper addresses the problem of universal
features associated with large scale ordering in superlattices. Suppose the
superlattice is built of atomically thick layers of two magnets, which taken
separately both order through second order phase transitions at two respective
bulk critical points $T_{c1}>T_{c2}$. Suppose also for simplicity that both
components are {\em ferro}-magnets and that the critical behavior at their bulk
Curie points is similar, i. e. both components belong to the same universality
class. The question then is: at what temperature $T_c$ will long range
ferromagnetic order appear in the superlattice and what critical behavior will
be observed at that temperature? One can reasonably expect that in the limit of
thick layers the answers to these questions will be largely universal,
depending only on the bulk critical properties of the two components and the
geometry of the superlattice; these universal aspects of the problem represent
the subject of this paper.

We will consider the simplest superlattice geometry (Fig. 1) constructed of two
elementary building blocks: slabs, or layers, of two ferromagnets, 1 and 2, of
finite thickness $L_1$ and $L_2$, respectively. The slabs are stacked
periodically in the $z$-direction, so that $L=L_1+L_2$ is the period. The
system is homogeneous in the remaining $d'=d-1$ dimensions.  While the
dimensionality $d=3$ (Fig. 1a) is naturally the most interesting one in view of
experimental applications, the planar, $d=2$, superlattice geometry can be
quite conceivably realized by cutting a thin film out of a three-dimensional
multilayered sample (Fig. 1b) or by deposition of a magnetic film on a
substrate, which is itself cut out of a superlattice. Extension of our
consideration to $d=2$ is important because of analytical tractability of
two-dimensional models. Being interested in the long wave length aspects of the
problem, we will consider the limit of atomically thick slabs: $L_1\gg a_1,\;
L_2\gg a_2$, where $a_{1,2}$ are the thicknesses of elementary, molecular
layers of the two components, respectively; correspondingly, the interfaces
separating two subsequents slabs do not have to be atomically smooth.
Naturally, universality extends applicability of the results obtained below
beyond this particular magnetic model; among other interesting applications
are a superfluid in a periodic matrix and a superfluid film deposited on a
periodic substrate.

Universal critical phenomena related to three different geometrical elements
present in a superlattice have been extensively studied theoretically and, to a
variable extent, experimentally. First, at a {\em single} interface separating
{\em half-infinite} near-critical, say $|t_1|\equiv |T-T_{c1}|/T_{c1}\ll 1$ and
disordered, $|t_2|\equiv |T-T_{c2}|/T_{c2}=O(1)$, magnets, one expects
{\em ordinary surface critical} phenomena to take place at the near-critical
side of the interface. It has been established theoretically
\cite{binder,diehl} that the critical exponents and amplitudes
characterizing surface values of various observables are quite different from
those in the bulk, as well as from the $d'$-dimensional ones. However,
experimental confirmation of those theoretical findings has turned out to be
difficult. Note that near $T_{c2}$, where the other half
of the single interface model goes through the criticality, the first one is
already ordered and imposes a magnetic field acting at the surface of the
second component. Therefore, one expects {\em normal surface critical} behavior
 \cite{LevFish,normal} to take place at the interface near $T_{c2}$.

Second, near the bulk critical temperature $T_{cj}$ of the $j$-th, $j=1,2$,
component a thick, $L_j\gg a_j$, {\em single} slab of that component exhibits a
dimensional crossover from the critical behavior characteristic of bulk,
$d$-dimensional, samples to the $d'=d-1$-dimensional critical behavior, as
observed in thin films of the same material \cite{bulk_to_film}. We will refer
to this crossover as {\em bulk-to-film} below; it has been observed
experimentally \cite{bulk_to_film_exp}. Finally, in a system of {\em thin}
layers connected to each other by very {\em weak} interlayer bonds, near the
critical temperature $T'_c$ of a single layer one expects a crossover from
$d'$-dimensional to the bulk, $d$-dimensional behavior, which we will call
{\em film-to-bulk} crossover below. This type of dimensional crossover has
been well experimentally documented in a variety of magnetic
\cite{Miedema de Jongh} and non-magnetic layered systems \cite{highTC}.

One source of motivation for studying critical phenomena in multilayers is that
the significantly increased surface-to-volume ratio may improve chances of
observing features characteristic of the $d=2$-dimensional and, especially,
the elusive $d=3$-surface critical behavior with
respect to experiments involving single thin film or semi-bulk samples.
However,
finite separation between interfaces in a superlattice leads to correlations
between fluctuations in different layers and at different interfaces. One
therefore expects
surface and two-dimensional scaling to be limited to a certain range of length
scales,
while eventually cooperative phenomena involving fluctuations on scales larger
than the period of the superlattice will dominate the criticality.
A crucial insight comes from the exact solution available for a ferromagnetic
$d=2$-Ising superlattice, as modelled by a planar Ising model
composed of alternating strips of two components \cite{Fish_Ferd}. Recent
analysis \cite{onsager} demonstrated that the anomalous changes in the critical
amplitudes and other interesting features of that model can be simply explained
by {\em reentrant dimensional crossover}: as the temperature approaches the
critical temperature of the superlattice the long wave length properties of the
system are successively dominated by fluctuations characteristic of the
the $d=2$-Ising, then $d=1$-Ising, and then again $d=2$-Ising critical
behavior.

Here we generalize and explore this picture by constructing and analyzing a
renormalization group (RG) flow which provides a scaling description of
criticality in
experimentally interesting three dimensional superlattices as well as in the
Heizenberg (isotropic) and $XY$ (easy plane) universality classes. In the
proposed
scenario as the scaled temperature $t=(T-T_c)/T_c$, or $t_1=(T-T_{c1})/T_{c1}$,
decreases, the critical behavior of the superlattice displays the features
of the three simpler systems listed above (Fig. 2): first, as $T_{c1}$ is
approached,
the original superlattice ((i) in Fig. 2) can be approximated by sequence of
thick
layers of the first component weakly coupled through the 2-layers ((ii) in Fig.
2).
At larger values of $t$ this coupling can be neglected and the layers
display regular bulk critical behavior
with the ordinary surface critical behavior at the interfaces. Then, as the
correlation length in the first component becomes comparable to the thickness
$L_1$, each slab of the first component exhibits direct bulk-to-film,
$d\rightarrow d'$-dimensional crossover. At still smaller values of $t$, the
thickness of these slabs is irrelevant and the system becomes equivalent to the
third paradigm mentioned above: a weakly coupled layered system ((iii) in Fig.
3).
The reverse, film-to-bulk crossover finally takes the system back to the
uniform, bulk $d$-dimensional behavior, characterized however by high degree
of anisotropy ((iv) in Fig. 2). This reentrant crossover behavior leads to
dramatic
changes in the critical amplitudes, as shown in Fig. 3 for the $d=3$-Ising
class: the critical exponent characterizing divergence of a physical quantity,
for which the specific heat per unit volume $C$ has been chosen in Fig. 3,
changes two times, as $t$ goes to zero, between the values $\alpha$ and
$\alpha^{\prime}$ characteristic of the $d=3$ and $d^{\prime}=2$-dimensional
critical behavior, respectively. The relative widths $\tau_D$ (subscript $D$
denotes {\em direct}, bulk-to-film crossover here) and $\tau_R$ (subscript $R$
stands for the {\em reverse}, film-to-bulk crossover) of the temperature
domains where the planar Ising and the reentrant bulk critical behaviors are
observed will be calculated below; they provide convenient scaling combinations
in terms of which a proper description of criticality in superlattices is
achieved.

Two further points have to be added to this preview of the paper. First, a
crucial element of our qualitative picture is that different layers of the
first component remain effectively decoupled while the bulk-to-film crossover
(from (ii) to (iii) in Fig. 2)
happens {\em within} them. This turns out to be guarantied by the fact that
the subsequent 1-slabs are coupled through the {\em surface} spins whose
correlations are much weaker than of those in the bulk. Thus the surface
critical behavior, which is not directly seen in temperature dependencies of
bulk quantities, like that shown in Fig. 3, appears crucial to the nature of
criticality in the superlattice; we will show that the values of surface
critical exponents can thus be extracted from analysis of the dependence of
the reentrant width $\tau_R$ on the thickness of the 2-slabs, $L_2$.

Second, in the case of the $d=3$ Ising superlattice the $d'=d-1$-dimensional
model, describing a single film of the first component, has a regular finite
temperature phase transition. Consequently, in the thick layer limit the
critical temperature $T_c$ of the superlattice is only slightly depressed from
$T_{c1}$ and the critical fluctuations are essentially localized within the
1-layers. However, in all other cases, i.e. $d=3$ Heizenberg and $XY$-models
and all models in $d=2$, the
corresponding $d'$-dimensional universality classes lack true ferromagnetic
long range order at finite temperatures. The interlayer coupling through the
layers of the weaker, second component thus are crucial to existence of long
range order. As a result, the critical temperature in these models is strongly
shifted towards $T_{c2}$ and the critical domain, in which large scale
fluctuations are active, is extended to the temperature range of the order of
$T_{c1}-T_{c2}$.

Finally, a few remarks about the method of this paper. We focus here on the
$n$-component vector spin systems with short range interactions. As usual,
Ising, $XY$, and Heizenberg universality classes refer to the $n=1$, or easy
axis, $n=2$, or easy plane, and $n=3$, or isotropic, magnets respectively.
The most important omission of this model, apart from restriction to the
simplest type of the order parameter, is the absence of long-range dipolar
forces; it seems however natural to make the first step within the simpler
realm of local models. The actual method employed here is to combine the well
known linear RG flows at the fixed points representing the three paradigms
discussed above (Fig. 2), into the simplest {\em global} RG flow consistent
with this
linear behavior (Fig. 4). The resulting scaling theory then passes the test of
comparison to the exact results available for the planar Ising model.

The outline of the paper is the following: in the next section we start by
taking the conceptually simplest case of the three-dimensional Ising
superlattice. The renormalization group flow will be proposed and scaling
forms derived. In section III we consider the planar Ising case which requires
extension of the previous consideration to the case of the zero-temperature
critical point describing the $d'=1$ dimensional Ising model. The results will
then be shown to agree with the exact forms available for this case
\cite{onsager}. The extension of the method developed in section III will be
used in the following section, where the Heizenberg and the $XY$ universality
classes are considered. The results are summarized in section V.

\section{Three-dimensional Ising superlattice}

We start now by focusing on the $d=3$-Ising universality class. Its most
important feature is that a single thick slab of the first component is fully
capable of ordering at a finite temperature $T'_{c1}$, which continuously
approaches the bulk transition temperature $T_{c1}$, as the thickness
$L_1\rightarrow\infty$. We thus expect that the criticality in a
three-dimensional Ising superlattice happens close to $T_{c1}$ and that the
relevant fluctuations are localized within the 1-slabs. Much of what will be
said below will apply however to other systems, so we will keep using general
$d$ and $d'=d-1$, as well as the general standard notation for the critical
exponents in this section.

The renormalization group procedure which we use here has been extensively
discussed in \cite{diehl} for the half-infinite models bounded by one free
surface, and applied to the superlattice geometry in \cite{onsager}. It is
based on subsequent reduction, $\Lambda'\rightarrow\Lambda'e^{-l}$, of the
upper cut-off $\Lambda'$ imposed on the $d'$-dimensional wave numbers
${\bf k'}$ characterizing the spatial variation of fluctuations in the
directions parallel to the layers. Since the superlattice geometry is uniform
in this plane, expansion in plane waves $\exp(i{\bf k'x'})$ can be used as a
basis for a perturbative RG of the type discussed in
\cite{diehl,HoughtonWegner}. The details of the procedure will not be
important to us here: all we need is the {\em existence} of an exact RG
transformation; then according to the general principles
\cite{HoughtonWegner,stellenbosch} the linearized RG flow in the vicinity of
the fixed points should be independent of the specific implementation. Note
that there is no need to do anything about the cut-off in the $z$-direction
(in fact one does not need such cut-off at all): elimination of fast modes in
the directions parallel to the layers automatically {\em induces}
coarse-graining in the $z$-direction \cite{diehl,onsager,frg}.

We will now follow the RG flow probing fluctuations on larger and larger
length scales parallel to the layers. The two crucial length scales, to which
the current (running) length scale has to be compared, are the {\em bulk}
correlation lengths $\xi_1(T)$ and $\xi_2(T)$, which can be (at least in
principle) measured at the given temperature $T$ in independent bulk samples
of the two components.
Near $T_{c1}$ fluctuations in the second component are confined to scales
smaller than the bulk correlation length $\xi_2(T)\approx \xi_2(T_{c1})$,
which remains finite while $\xi_1$ diverges.
Decreasing the upper momentum cutoff to $\xi_2^{-1}$ essentially eliminates
fluctuations in the 2-layers. On larger scales the 2-layers are adequately
described by the Ornstein-Zernike spin density functional corresponding to the
Gaussian RG fixed point describing the low-temperature phase of the second
component. As usual, Gaussian degrees of freedom can be integrated out, the
integration being equivalent to minimization of the Ornstein-Zernike
functional. Such minimization is performed in the Appendix. It results in an
effective interaction \cite{wetting}
\begin{equation}
J_0e^{-L_2/\xi_2}\int {\bf s}_n^T(x){\bf s}_{n+1}^B(x) d^{d-1}x' \label{int}
\end{equation}
between the top surface spins ${\bf s}_n^T(x)$ of the
$n$-th layer and the bottom surface spins ${\bf s}_{n+1}^B(x)$ of
the $n+1$-th layer of the first component. At this point the thickness
$L_2$ gets absorbed into the bare coupling constant $J_i=J_0\exp(-L_2/\xi_2)$.
A major simplification has occurred after this initial crossover: the RG flow
has taken the original superlattice system to the first fixed point, $FP_0$,
equivalent to a sample of the first component containing a periodic sequence
of defect (hyper-) planes, characterized by
very weak vertical bonds $J_i$ ((ii) in Fig. 2). Note that all physics of the
second
component, as well as all details of the microscopic implementation of the i
interface between the subsequent layers, have been absorbed into two
parameters:
 the directly measurable correlation length $\xi_2$ and the nonuniversal
amplitude $J_0$.

A necessary condition for the reentrant dimensional crossover, described in
the Introduction, is the smallness of the initial, bare value of the coupling
in the temperature range around $T_c\approx T_{c1}$:
\begin{equation}
J_i=J_0\exp[-L_2/\xi_2(T_{c1})]\ll k_BT.
\label{condition1}
\end{equation}
Under this condition, we can consider the coupling (\ref{int}) as a weak
perturbation with respect to the reference system consisting of
{\em noninteracting} layers of the first component. Further, for the values of
the RG parameter $l$ such that $\xi_2\ll\Lambda'\ll L_1$, typical fluctuations
are correlated only over lengths much shorter than $L_1$, implying that the
top and the bottom surfaces of each 1-layer are uncoupled. Provided the
correlation length $\xi_1$ of the first component has not been encountered yet,
the system appears to be at the critical fixed point $CFP_s$ describing
independent bulk critical behavior within the 1-layers coupled to the ordinary
surface critical behavior at the noninteracting interfaces. Note that this
scenario implies another condition,
\begin{equation}
L_1\gg\xi_2(T_{c1}).
\label{condition2}
\end{equation}
Both conditions, (\ref{condition1},\ref{condition2}), require the layers to be
thick.
In the case of two different materials, $T_{c1}-T_{c2}=O(1)$, so that
$\xi_2(T_{c1})=O(a_2)$ and both inequalities are satisfied as soon as
$L_1, L_2\gg a_2$. The situation becomes less trivial in the case of a
superlattice created by a weak periodic modulation of the properties of an
originally uniform sample, then one has to be sure that the period of
modulation is large enough to offset the smallness of the difference between
the bulk Curie temperatures implying a relatively large $\xi_2(T_{c1})\propto
(T_{c1}-T_{c2})^{-\nu}$, where $\nu$ is the standard correlation length
exponent in $d$ dimensions.

The crossover is now described as an RG-flow between three critical
fixed points (Fig. 4): $CFP_s$ describing a sequence of uncorrelated
{\em thick} layers of the first component ((ii) in Fig. 2), with
extensive properties dominated by the $d$-dimensional bulk critical behavior
taking place inside the layers, while the associated ordinary surface scaling
describes the observables localized at the surface.
{}From $CFP$ the system flows to $CFP'$ describing the
$d'=(d-1)$-dimensional criticality in a system of uncorrelated {\em thin}
films of the first component ((iii) in Fig. 2), and, finally,
to $CFP_b$ describing uniform $d$-dimensional bulk behavior ((iv) in Fig. 2).
The flow between the fixed points is driven by two scaling fields (cf. Fig. 4),
the inverse thickness of the 1-layers, $L_1^{-1}$, playing the role of a long
wave
length (infrared) cutoff in the $z$-direction (cf. \cite{StevensO'Connor}),
and the interlayer coupling strength $J_i$. Yet another one, scaled temperature
field $t_1=(T-T_{c1})/T_{c1}$ controls the departure from the critical
manifold, $t_{1c}(L_1, J_i)$, containing the flow attracted by $CFP_b$,
towards the massive, high- and low-temperature fixed points.
Instead of the scaled temperature field $t_1$ characterizing the bulk
criticality of the first component, one can use $t=(T-T_c)/T_c$ defined
relative to the {\em observed} critical temperature $T_c(L_1, J_i)$ of the
superlattice. As we will see below, the scaling forms look simpler when
expressed via $t$, but lack any information about the {\em shift} in the
critical temperature with respect to $T_{c1}$. In the RG approach the scaling
fields become functions of the logarithmic length scale $l$. While the
explicit calculation of the crossover scaling
functions requires application of non-perturbative methods
such as the one used in \cite{onsager}, the leading singularities are
determined
by the RG flow in close vicinities of the fixed points.

Specifically, at $CFP_s$ the linearized RG flow equations are
\begin{eqnarray}
dt_1/dl=(1/\nu)t_1,\label{tD}\\
d(1/L_1)/dl=1/L_1,\label{LD}\\
dJ_i/dl=(d'-2\omega_1)J_i=(\gamma_{11}/\nu)J_i.\label{JD}
\end{eqnarray}
Here $\nu$ is the standard correlation length exponent of the bulk
$d$-dimensional universality class. The second equation (\ref{LD}) reflects
decrease in $L_1$ expressed in units of the running inverse length scale
$\Lambda'$; the equality of the RG eigenvalue in (\ref{LD}) to one reflects
asymptotic isotropy of the bulk critical behavior described by $CFP_s$.
The RG eigenvalue for $J_i$ is read from (1). Because the spins entering (1)
are located at the
surfaces, the scaling dimension of the spin density $\omega_1$ determining the
RG eigenvalue of $J_i$ is that characteristic of the {\em surface} spin
density. Correspondingly $\gamma_{11}$
is the exponent characterizing the susceptibility of the {\em surface} spins to
a perturbation by a {\em surface} field \cite{diehl}. The crucial point now is
that since the surface
spins are correlated more weakly than the bulk ones, the surface susceptibility
usually does not diverge (see the estimates of surface critical exponents in
\cite{diehl}): $\gamma_{11}\leq 0$. Consequently $J_i$ is irrelevant or
marginal at the first encounter with $CFP$. Thus starting with a small
coupling $J_i$, we are guarantied that it remains small until the system
arrives at $CFP'$. Note that this picture implies a seemingly paradoxical
prediction: a single layer of weak bonds cutting through a bulk sample
effectively decouples the two halves, with two {\em independent} ordinary
surface fluctuations developing at the two sides of it.

The two other scaling fields, $t_1$ and $L_1^{-1}$ are relevant at $CFP_s$.
As $L_1(l)$ is positive definite, it is convenient to exclude $l$ from the
equations (\ref{tD},\ref{LD}), rewriting them as
\begin{eqnarray}
d\ln(t_1)/d\ln(L_1)=-\nu^{-1},\label{tL}\\
d\ln(J_i)/d\ln(L_1)=-\gamma_{11}\nu^{-1}.\label{JL}
\end{eqnarray}
The crossover to the $d'$-dimensional fixed point $CFP'$ occurs when
$L_1(l)$ decreases to the order of the microscopic length scale $a_1$. As a
result of this first crossover (indicated by subscript $D$, for ``direct,''
below) the scaling field $1/L_1$ becomes irrelevant and disappears from the
consideration, while the other two are renormalized to
\begin{eqnarray}
t_{1D}=t_1/\tau_D \label{t1D}\\
J_{iD}=J_i\tau_D^{-\gamma_{11}}.
\end{eqnarray}
Here the temperature rescaling factor
\begin{equation}
\tau_D\approx (a1/L_1)^{1/\nu} \label{tauD}
\end{equation}
conveniently characterizes the first crossover. We will reduce the number of
nonuniversal parameters by defining $\tau_D$ not through $a_1$, but rather
through the critical amplitude $X_1=O(a_1)$ characterizing the divergence of
the bulk correlation length $\xi_1=X_1(-t_1)^{-\nu}$ at the low-temperature
side of the bulk critical point $T_{c1}$:
\begin{equation}
\tau_D= (X_1/L_1)^{1/\nu}.
\end{equation}
In the RG formalism \cite{stellenbosch} the correlation length $\xi_1$
characterizes the crossover from the critical to noncritical RG fixed points
at nonzero values of $t_1$. Therefore $\tau_D$ determines the width of the
temperature domain around $T_{c1}$ in which the dimensional crossover
$CFP_s\rightarrow CFP'$ actually occurs: for
$t_1>\tau_D$ the $t_1$-field becomes large and drives the system away from the
critical manifold before $1/L_1$ grows large; as a result the system never
reaches  $CFP'$.

If, on the other hand, $t_1\lesssim\tau_D$ the system arrives at $CFP'$ with
the values of the two relevant fields estimated by
$t_1\approx t_{1D}, \; J_i\approx J_{iD}$. In the
absence of $J_i$ (which is small at this stage) $t_{1}$ completely determines
the flow. The critical separatrix going into $CFP'$ corresponds to a certain
initial value $t_{1D}=t'_c=O(1)$. One expects $t'_c<0$, as finite thickness
suppresses ordering. In fact, due to our definition of $\tau_D$ via the bulk
correlation length $\xi_1$, the parameter $t'_c$ is universal: criticality in
a {\em free} film of the first component of thickness $L_1$ happens at a
temperature $T'_c(L_1)$ at which the ratio $L_1/\xi_1$ takes a universal value
\begin{equation}
L_1/\xi_1(T'_c)=(-t'_c)^{\nu}\approx 2.89,
\end{equation}
where the numerical estimate has been obtained by series expansion methods
\cite{Nakanishi_Fisher}. Expanding around the separatrix one obtains
\begin{eqnarray}
dt_{1}/dl=(1/\nu')(t_{1}-t'_c)\\
dJ_{i}/dl=(d'-2\omega')J_{i}=(\gamma'/\nu')J_{i},
\end{eqnarray}
where the prime marks exponents related to the $d'$-dimensional criticality.
These equations are to be solved with the initial data given by
$t_1=t_{1D},\; J_i=J_{iD}$. As a result of the first
crossover the surface spins in (\ref{int}) are strongly correlated with the
rest of the corresponding layer, thus becoming $d'$-dimensional in nature.
Correspondingly, the scaling dimension of the spin density $\omega_1$ in the
RG flow equation for $J_i$ is changed from the {\em ordinary surface} value
$\omega_1$ to the $d'$-dimensional $\omega'$.
Correspondingly nonpositive $\gamma_{11}$ is changed to positive
$\gamma'$, so that $J_i$ is a relevant perturbation at $CFP'$ driving the
second
crossover back to $CFP$. The following analysis of the reverse crossover
(as indicated by subscript $R$ below) essentially repeats the one performed
for $CFP_s\rightarrow CFP'$ above: we divide one of the two linear RG
equations by another to obtain (the coupling strength $J_i$ is positive
definite)
\begin{equation}
d\ln(t_{1}-t_{ac})/d\ln J_{i}=1/\gamma'.
\end{equation}
A nonuniversal amplitude $J_0^*=O(k_BTD^{-d'})$ is defined so that at
$J_i(l)\gtrsim J_0^*$ the $d'$-dimensional hyperplanes become strongly coupled
and the system crosses over to the uniform bulk behavior at $CFP_b$. As this
happens, $t_1$ is rescaled by the factor $\tau_R=(J_{iD}/J_0^*)^{1/\gamma'}$,
i. e.
\begin{equation}
t_1(l)=t_{1R}\equiv (t_{1D}-t'_c)/\tau_R,
\label{t1R}
\end{equation}
where (recall that $\gamma_{11}\leq 0$)
\begin{equation}
\tau_R=(J_0/J_0^*)^{1/\gamma'} (L_1/X_1)^{-|\gamma_{11}|/\gamma'\nu}
\exp(-L_2/\gamma'\xi_2).
\label{tauR}
\end{equation}
Just as previously, for $|t_{1R}|\gg 1$, i. e. $|t_{1D}-t'_c|\gg \tau_R$,
the system flows away from the critical
separatrix before the reentrant crossover to $CFP$ can take place. Hence
$\tau_R$ measures in units of $t_{1D}$ the width of the temperature domain in
which the reentrant scaling can be observed; in the original units of $t_1$ the
width is given then by the product $\tau_D\tau_R$. Within the domain
$t_{1R}\lesssim 1$ the system flows into the close vicinity of $CFP_b$. The
critical separatrix is again defined by a certain (this time positive:
interlayer bonds enhance ordering) value of $t_{1R}=t_{cR}=O(1)$. In fact, as
at this point we have no other scale to measure the interlayer coupling
amplitude $J_0$, we can fix the arbitrary constant $J_0^*$ by setting
$t_{cR}=1$. Having grown
to the order of $J_0^*$ the interlayer coupling amplitude $J_i$ is absorbed
into the rescaled value of the temperature field $t_1(l)=t_{1R}$ and becomes
irrelevant. The reentrant $CFP_b$ is a regular critical fixed point with
temperature (in the absence of magnetic field) being the only relevant
perturbation measuring the deviation from the critical separatrix (Fig. 4).
The evolution of the latter is again described by the RG flow equation
(\ref{tD}) which has to be solved with the initial condition
\begin{equation}
t_1(l)=t_{1R}-1=(t_1/\tau_D-t'_c)/\tau_R-1.
\label{t1c}
\end{equation}

Having established the principle features of the RG flow we are in a position
now to develop a scaling description of the observable quantities.
The two temperature scales $\tau_D$ and $\tau_R$ defined by (2) and (3),
together with the original value of the scaled temperature of the first
component, $t_1$, conveniently parametrize the scaling functions. If the bulk
properties of the two components are known, then the only new nonuniversal
parameter appearing in our description of the superlattice is the
dimensionless ratio $J_0/J_0^*$, essentially characterizing the strength of
coupling between the spins across the interface between the two components
(see the Appendix for more precise definition). The standard two-scale factor
universality \cite{HohenbergPrivman} of the bulk critical points is thus
extended to what may be called $2+2$-factor universality: all scaling
functions of a superlattice are universal apart from the two independent bulk
critical amplitudes of the first component plus the amplitude $J_0/J_0^*$
characterizing the interface and the bulk correlation length of the second
component $\xi_2(T_{c1})$. In fact, the two additional parameters enter the
description via the initial value of the effective coupling
$J_i(l=0)=J_0\exp[L_2/\xi_2(T_{c1})]$. Thus, if one were not interested in the
(singular) dependence of the critical behavior of the superlattice on the
thickness $L_2$, only one extra amplitude, $J_i$, would have to be added to
complete the description of critical behavior in a superlattice. However,
since for large $L_2$ the amplitude $J_i$ is going to be anomalously small,
and since $L_2$ is easily measured, we prefer to split $J_i$ into the
nonsingular amplitude $J_0$ and the singular exponential factor
$\exp(-L_2/\xi_2)$, adding the bulk correlation length $\xi_2(T_{c1})$ to the
list of empirical parameters.

We start by analyzing the shift in the critical temperature. Unfolding back
the two renormalizations of the temperature field, one obtains from
(\ref{tD},\ref{t1c})
\begin{equation}
T_c=T_{c1}[1+\tau_D (t'_c+\tau_R)]=T'_c(L_1)+T_{c1}\tau_D\tau_R.
\end{equation}
The shift consists of a larger, $O(L_1^{-1/\nu})$, shift to lower temperatures,
slightly corrected by a much smaller, $\;O[\exp(-L_2/\xi_2)]$, shift in the
opposite direction. The last expression represents the latter as a shift with
respect to $T'_c(L_1)=T_{c1}[1+\tau_D t'_c]$, the critical temperature in a
film of the first component of thickness $L_1$ with free boundaries. If this
latter temperature is known, then the observed shift towards the higher
temperatures, $T_c-T'_c(L_1)$, can be used to estimate the unknown amplitude
$J_0/J_0^*$ entering the definition of $\tau_R$ (\ref{tauR}).

Let us consider now an extensive observable, say the specific heat per unit
volume $C$ characterized by critical exponents $\alpha$ and $\alpha'$ in
dimensions $d$ and $d'$ correspondingly. Since the surface-to-volume ratio in
a superlattice vanishes as $L_{1,2}\rightarrow\infty$, the observed signal at
the first stage of the crossover is dominated by the interior of the 1-layers:
as far as extensive properties are concerned $CFP_s$ is equivalent to $CFP_b$
and the RG flow is indeed reentrant. This flow can be represented by the
scaling form
\begin{equation}
C(t_1)=(L_1/L)A_{1-}|t_1|^{-\alpha}{\cal A}_1(t_{1D},t_{1R}),
\label{A1}
\end{equation}
where $t_{1D},\; t_{1R}$ are given in (\ref{t1D},\ref{t1R}), and $A_{1-}$ is
the bulk critical amplitude of the first component on the low-temperature side
of the criticality. The factor $(L_1/L)A_{1-}$ represents the amplitude of the
contribution to $C$ from the interior of the 1-layers at the first stage of
the RG flow (governed by $CFP_s$) below $T_{c1}$; the choice of $A_{1-}$ i
instead of $A_{1+}$ seems natural in view of $T_c<T_{c1}$.
The scaling function ${\cal A}_1$ is completely {\em universal}. It has the
following
limits: it approaches 1 at $t_{1D}\rightarrow -\infty$ and the universal
amplitude ratio $A_{1+}/A_{1-}$ at $t_{1D}\rightarrow +\infty$; at
$t_{1D}\rightarrow 0$ it develops a singularity
${\cal A}_1\propto |t_{1D}|^{\alpha}$ which cancels the one generated
by the $|t|^{-\alpha}$ factor in (\ref{A1}). At $|t_{1D}-t'_c|\ll 1$ one has
${\cal A}_1\approx |t_{1D}-t'_c|^{-\alpha'}{\cal C}_1(t_{1R})$. The asymptotic
behavior of the new scaling function ${\cal C}_1$ follows the same logic as the
one employed above for ${\cal A}_1$: it takes finite limits at
$t_{1R}\rightarrow\pm\infty$, develops a singularity, ${\cal C}_1\propto
|t_{1R}|^{\alpha'}$ at $t_{1R}\rightarrow 0$ to compensate for the singular
prefactor, and, finally, diverges as ${\cal C}_1\propto |t_{1R}-1|^{-\alpha}$,
when $t_{1R}\rightarrow 1$.

As already mentioned above, because in the Ising universality class the
critical
temperature shifts satisfy $\tau_D\tau_R\ll\tau_D t'_c\ll 1$ a simpler
scaling form is achieved in terms of the scaled temperature
$t=(T-T_c)/T_c=t_1-\tau_D(t'_c+\tau_R)$ shifted to the {\em observed}
critical temperature $T_c(L_1,J_i)$. Defining $t_D=t/\tau_D,\; t_R=t_D/\tau_R$
one can
write
\begin{equation}
C=(L_1/L)A_{1-}|t|^{-x}{\cal A}(t_D, t_R),
\end{equation}
where the universal scaling function ${\cal A}$ takes the following limits:
\begin{eqnarray}
{\cal A}\approx {\cal A}_{1+}/{\cal A}_{1-}, \; {\rm at}\;  1\ll t_D\ll t_R,\\
{\cal A}\approx 1, \; {\rm at}\;   t_R\ll t_D\ll -1,\\
{\cal A}\approx C_{D\pm}t^{-\alpha}t_D^{\alpha-\alpha'}, \; {\rm at}\;
|t_D|\ll 1\ll |t_R|,\\
{\cal A}\approx C_{R\pm}t^{-\alpha}t_D^{\alpha-\alpha'}t_R^{\alpha'-\alpha}, \;
{\rm at}\;   |t_D|\ll |t_R|\ll 1.
\label{A}
\end{eqnarray}
This form has a simple graphical interpretation given by the double logarithmic
plot of Fig. 3. The continuity of $C(t)$ requires that both universal amplitude
ratios
$C_{D\pm},\; C_{R\pm}$ (remember that the scaling function has been normalized
by the
bulk amplitude $A_{1-}$) are numbers of the order of unity. The last asymptotic
expression in (\ref{A}) shows that while the ultimate divergence
$A=A_{\pm}|t|^{-\alpha}$
is characterized by the bulk exponent $\alpha$, the critical amplitudes
\begin{equation}
A_{\pm}=(L_1/L)A_{1\pm}C_{R\pm}\tau_R^{\alpha-\alpha'}
\label{Cratio}
\end{equation}
are shifted on the logarithmic scale from the bulk amplitudes $A_{1\pm}$ of the
first
component by $(\alpha'-\alpha)\ln(1/\tau_R)$ (see Fig.3), i.e.
\begin{equation}
\frac{\ln(A_{\pm}/A_{1\pm})}{\alpha'-\alpha}=\frac{1}{\gamma'}[L_2/\xi_2(T_c)+
(-\gamma_{11}/\nu)\ln(L_1)]+O(1).
\label{logCratio}
\end{equation}

More information about the critical state of a superlattice governed by $CFP_b$
can be
obtained by studying the anisotropy of correlation functions. The basic ratio
$X=\xi_{\perp}(T)/\xi_{\parallel}(T)$, of the critical amplitudes of
correlation lengths
across and along the layers, is given by expressions identical to
(\ref{Cratio}),
(\ref{logCratio}), where one has to use anisotropic correlation length
exponents:
$\nu_{\perp}=\nu_{\parallel}=\nu$ at $CFP_s,\; CFP_b$, but $\nu_{\perp}=0,\;
\nu_{\parallel}=\nu'$ at $CFP'$, since at $CFP'$ the correlations grow only in
the
direction along the layers. The result is:
\begin{equation}
X\sim \tau_R^{\nu'}.
\end{equation}
Similarly, at $T_c$, the amplitude of the spin-spin correlation function,
$G_{\perp}(z)\propto z^{-2\omega}$ at $z\gg L$ is suppressed compared to that
of
$G_{\parallel}(x)\propto x^{-2\omega}$ by an amplitude ratio \begin{equation}
\Omega\equiv G_{\perp}/G_{\parallel}\sim X^{2\omega}\sim\tau_R^{2\beta'}.
\end{equation}
One thus arrives at the large-scale description of the critical state of the
superlattice
as a highly anisotropic realization of the bulk universality class of the first
component
((iv) in Fig. 2). In fact, following the standard ideas of scaling
\cite{FisherNobel} one can
relate all anomalous changes in the critical amplitudes (\ref{Cratio}),
(\ref{logCratio})
to the anisotropy $X$ of the basic length scales.

\section{Planar Ising superlattice: Extension to the case of zero-temperature
criticality
in $d'=1$ dimensions and comparison to exact results}

Despite the generality of the above consideration, of all the $O(n)$ spin
lattice models
in three and two dimensions only the $d=3,\; n=1$-superlattice behaves strictly
according
to this scenario, as the Ising model has regular finite-temperature critical
fixed points
without marginal operators in both $d=3$ and $d'=2$. The exactly solvable
problem of a
$d=2$-Ising superlattice requires certain modifications because $CFP'$
describing the one-dimensional Ising model is a zero-temperature fixed point.
This results \cite{onsager} in a {\em large} shift in the critical temperature,
$T_{c1}-T_{c}=O(1)$. More precisely, in the interesting case of a {\em thick
layer}
superlattice \cite{onsager}, criticality happens when the correlation length
$\xi_1$,
characterizing the low-temperature phase of the first component, is much
smaller than the
width of the 1-layers, $L_1$. This means that the RG flow in this case comes to
$CFP'$
not from $CFP_s$, but from the low-temperature fixed point $LTFP$ describing
the ordered
phases of the first component. Nevertheless, many results can be formulated in
terms of
just one nonuniversal parameter characterizing $LTFP$: the linear energy
$\Sigma_1$ of a
domain wall separating spin-up and spin-down phases of the first component.

Let us start our consideration from a close vicinity of $T_{c1}$, where
$\xi_1\gg L_1$
and the crossover $CFP_s\rightarrow CFP'$ proceeds directly, without
encountering $LTFP$.
After leaving $CFP_s$ the scaled
temperature $t_1$ is renormalized to $t_{1D}=t_1(L_1/X_{1-})^{1/\nu}$. Since
$\nu=1$ in
the planar Ising universality class, below but close to $T_{c1}$ one can write
$t_{1D}=-L_1/\xi_1$ (note that the normalization $|t_1|=1/\xi_1$, making
$X_{1-}=X_{1+}=1$ has been adopted in \cite{onsager}). Further, $\gamma_{11}=0$
in this
universality class \cite{LevFish}, so that the coupling
$J_i\propto\exp(-L_2/\xi_2)$ is not renormalized near $CFP_s$ to the linear
order in
$J_i$. While one should expect logarithmic corrections to $J_i$ for this
marginal case,
those do not seem to affect the essential features of the exact solution. As
discussed in
\cite{onsager}, at $t_1\ll 1$
\begin{equation}
|t_{1D}|=L_1/\xi_1=2L_1\Sigma_1/k_BT\equiv 2J'.
\label{xi-sigma}
\end{equation}
The last identity, in which $\Sigma_1(T)$ is the linear free energy of a domain
wall
between the two low-temperature phases of the first component, represents the
dimensionless coarse-grained spin-spin coupling $J'$ of the $d=1$-Ising model
onto which
a strip of the first component is mapped after the direct, bulk-to-film
crossover.
At $CFP'$ this spin-spin coupling transforms according to $dJ'/dl=-1/2$, or
\begin{equation}
dt_1/dl=1
\end{equation}
with the critical separatrix corresponding to the extreme value
$t'_{c}=-\infty$,
as implied by $T'_c=0$ in the $d'=1$-dimensional Ising class. Note that the
fugacity
$\zeta=\exp(-2J')$, taking the initial value $\exp(t_{1D})$, behaves at $CFP'$
as a
regular thermal scaling field with the RG eigenvalue equal to one.
Consequently, the
criterion for $CFP'$-dominated behavior is $\zeta\ll 1$, or, equivalently,
$t_{1D}\ll -1$.

Meanwhile, magnetization scales trivially, $\omega'=0$, at $CFP'$. Thus the
eigenvalue of
$J_i$ is $d'-2\omega'=d=1$
and the corresponding RG equation at $CFP'$ can be
rewritten as $d\ln J_i/dl=1$. Dividing one RG equation by another we obtain
\begin{equation}
dt_{1}/d\ln J_i=1,
\end{equation}
from which one can construct a renormalized temperature
\begin{equation}
t_{1R}=t_{1D}-\ln(J_i/J_0^*)=L_2/\xi_2-L_1/\xi_1+O(1).
\end{equation}
It is convenient to define $J_0^*$ here in such way that
the criticality occurs at $t_{1R}=0$. Then, in full agreement with
\cite{onsager}, up to
corrections vanishing as $O(L_{1,2}^{-1})$,
the critical temperature $T_c$ of the superlattice can be obtained from the
equation
\begin{equation}
L_1/\xi_1(T_c)=L_2/\xi_2(T_c)\equiv g_c,
\end{equation}
where the last identity defines what may be called the {\em scaled thickness}
of the layers.

If this equation is satisfied at $g_c\ll 1$, the case called the {\em thin
layer} limit
in \cite{onsager}, then the flow indeed never encounters $LTFP$. However, in
this case
the initial value of the coupling $J_i\propto e^{-g_c}$ is not small, so that
the
criticality is reached before $t_{1D}=-g_c$ becomes large and negative so as to
display
any features of the $d'=1$ critical behavior.
The flow in fact never leaves $CFP$. The reentrant crossover does happen in the
opposite,
{\em thick layer} limit \cite{onsager}, $g_c\gg 1$, however in this case the
behavior of
the system on length scales between $\xi_1$ and $L_1$ is governed by the
low-temperature,
non-critical fixed point $LTFP$.

To describe criticality in this limit we start at scales larger than $\xi_1(T)$
well
below $T_c$. If the RG is implemented with rescaling the block spin by the
factor of the
area of the block (rather than the square root of the area leading to the
Gaussian fixed
point described by the Ornstein-Zernike functional \cite{parizi}), then at
these scales
fluctuations in the amplitude of magnetization are completely suppressed. The
only
allowed fluctuations are domain walls separating domains of two different spin
orientations. At these scales the walls look geometrically sharp and are
characterized
(neglecting lattice effects) by a single parameter: free energy per unit length
$\Sigma_1$.
As discussed in \cite{onsager}, a single strip of the first component becomes
identical
to a
$d=1$-Ising model: spins are fully correlated across the strip; a domain wall
cutting
across the strip has finite free energy $\Sigma_1L_1$ playing a role of an
effective
ferromagnetic coupling between neighboring blocks of size $\xi_1\times L_1$.
Characterization of the effective $d=1$-model is completed by specifying the
value of a
block spin, $m_1(T)L_1\xi_1$, where $m_1(T)$ is the bulk magnetization density
in the
first component.

The crossover from $CFP'$ to $CFP_b$ proceeds exactly as described in the
beginning of
this section, because the representation (\ref{xi-sigma}) of $t_{1D}$ via the
linear
surface energy $\Sigma_1$ is valid independently of whether the system has
arrived at
$CFP'$ from $CFP_s$ or from $LTFP$. After the reverse crossover the thermal
field is
renormalized to
\begin{equation}
t_{1R}=t_{1D}-\ln(J_i/J_0^*)=L_2/\xi_2-L_1\Sigma_1/k_BT+O(1),
\end{equation}
leading to the universal relation for the critical temperature
\begin{equation}
\frac {\Sigma_1(T_c) \xi_2(T_c)}{k_BT_c}=\frac{L_2}{L_1}
\label{Tc2D}
\end{equation}
holding up to corrections of order $O(a/L)$. Note that the relation
$\Sigma_1=k_BT/2\xi_1$ used in (\ref{xi-sigma}) is valid in the planar Ising
universality class up to corrections of order $O(a/\xi_1)$, which were
neglected in
the field-theoretical description of Ref.\cite{onsager}. The expression
(\ref{Tc2D})
however has a much wider range of validity: it requires only that the strips
are thick,
with no condition on the values of the correlation lengths in the two
components.

For an extensive observable like the specific heat per unit area we can write
now
\begin{equation}
C=\frac{1}{L} A'(T)\exp(-\tilde{\alpha}'t_{1D}){\cal A}(t_{1R}).
\label{2DIsing}
\end{equation}
Here the exponent $\tilde{\alpha}'$, equal to $-1$ in the case of the specific
heat,
and the critical amplitude $A'(T)$ characterize singularity
$C=A'\zeta^{-\tilde{\alpha}'}$
in the one-dimensional regime. The scaling function ${\cal A}$ has the
following limits:
it approaches 1 at $t_{1R}\gg 1$; diverges as
${\cal A}\approx {\cal C}_{\pm}t_{1R}^{-\alpha}$ at $t_{1R}\ll 1$ with the
amplitude
ratios ${\cal C}_{\pm}=O(1)$ being universal; an exponential singularity
${\cal A}\propto\exp(\tilde{\alpha}'t_{1R})$ cancels the singular prefactor in
(\ref{2DIsing}) at $t_{1R}\rightarrow -\infty$. The critical divergence of the
specific
heat,
\begin{equation}
C\approx \frac{1}{L} A'(T_c)\exp(-\tilde{\alpha}'g_c){\cal
C}_{\pm}t_{1R}^{-\alpha}=
At^{-\alpha},
\label{c2Dfull}
\end{equation}
is characterized by the amplitude
\begin{equation}
A\sim A'(T_c)\exp(-\tilde{\alpha}'g_c),
\label{A2D-1}
\end{equation}
where we have taken into account that $dt_{1R}/dt=O(L)$ cancels the $L^{-1}$
prefactor
in (\ref{c2Dfull}).

In order to complete the estimate of $A$ we have to know the amplitude
$A'(T_c)$.
The one-dimensional amplitudes of extensive observables (per unit {\em length}
of a layer)
can be generated from the well-known expression for the free energy density
$f'$ of a
one-dimensional chain at temperature $T$ and magnetic field $H$ (see
\cite{levfish1} and
references there)
\begin{equation}
f'/k_BT\sim\left ((Hm_1L_1)^2+(\zeta/\xi_1)^2\right )^{-1/2},
\label{f1D}
\end{equation}
where we have taken into account that $\xi_1$ plays the role of the effective
cutoff for
the one-dimensional behavior along the layers, and both $\xi_1$ and $m_1$ are
taken at
$T=T_c$. Differentiating with respect to the temperature $T$ we obtain (at
$T=T_c$ and
$H=0$) $A'(T)= g_c^2k_B/\xi_1$. Combined with (\ref{A2D-1}) this gives
\begin{equation}
A\sim k_B g_ce^{-g_c}\xi_1^{-2}.
\label{A2D-2}
\end{equation}
This agrees fully with the result of \cite{onsager}. Other amplitudes can be
similarly
estimated and agree with the exact solution.

An interesting feature of the form (\ref{2DIsing}) is that below $T_c$, at
$t_{1R}\ll -1$ the
exponential part of the scaling function ${\cal A}$ compensating for the (now
spurious)
singular prefactor, leads to
\begin{equation}
C\propto \exp(\tilde{\alpha}'L_2/\xi_2).
\end{equation}
This can now be recognized as pointing to the dominance of the fluctuations in
the
{\em second} component as $T_{c2}$ is approached: in fact, due to the self
duality of the
planar Ising model, a completely similar construction can be
developed starting from low-temperatures and focusing on the dual variables in
the
2-layers. Both scenarios lead to essentially the same results.

\section{Heizenberg and $XY$ superlattices}
The other $O(n)$ models can be analyzed similarly; only the most interesting
results will be briefly discussed below.

The case of the three-dimensional Heizenberg ($n=3$) model is very close to the
planar
Ising one,
as the $(d'=2, n=3)$-class is believed to be characterized by a
zero-temperature critical
point
with the effective coupling being marginally relevant \cite{Pokr_Patash}. This
effective
coupling in a single Heizenberg film of the first component is the
dimensionless
{\em helicity modulus} \cite{FisherPrivman} $\Gamma_1'$. In the low-temperature
phase of
the $d=3$-Heizenberg model the helicity modulus $\Gamma_1$ has the dimension of
inverse
length, effectively defining the correlation length $\xi_1\equiv\Gamma_1^{-1}$.
In the
thick layer limit the critical temperature $T_c$ of the superlattice is again
shifted
well below
$T_{c1}$ of the first component. Thus on scales smaller than $L_1$ the first
component is well ordered and the additive approximation $\Gamma'=\Gamma L_1\gg
1$ is
well justified \cite{FisherPrivman}. Here $\Gamma_1$ can be (at least in
principle)
measured independently in a bulk sample of the first component. Once the direct
crossover
to $CFP'$ has occurred, one can form $t_{1D}=k_BT/\Gamma'$ and use the well
known
$d=2$-Heizenberg model RG equation \cite{Pokr_Patash}
\begin{equation}
dt_{1D}/dl=1/2\pi,
\end{equation}
while the coupling $J_i$ is trivially additive
\begin{equation}
dJ_i/dl=d'J_i,
\label{additive}
\end{equation}
with $d'=2$ here. Combining these two equations we obtain, in complete analogy
with
(\ref{Tc2D}), a universal expression satisfied at the critical point of a
Heizenberg
superlattice:
\begin{equation}
\frac{\Gamma_1(T_c)\xi_2(T_c)}{k_BT_c}=\frac{L_2}{4\pi L_1}+O(L^{-1}).
\label{TcH3D}
\end{equation}
Quite analogously to the consideration given above for the planar Ising case
one arrives at the picture
of the critical temperature shift $T_{c1}-T_c=O(T_{c1}-T_{c2})$, large scale
fluctuations
between in the wide temperature range between $T_{c1}$ and
$T_{c2}$, and exponential dependence of critical amplitudes on
the thickness of the layers.

The case of a $XY$ ($n=2$) superlattice in $d=3$ is special because, on one
hand, the
$d'=2$-class is characterized by a finite-temperature
Kosterlitz-Thouless transition, so that we expect the shift in the
critical temperature to vanish in the limit of (atomically) thick layers. On
the other
hand, the spontaneous magnetization is zero in $d'=2$ at any finite
temperature. Thus the
coupling $J_i$ through the 2-layers is crucial in supporting long-range order
in a
three-dimensional superlattice and we
expect the observed spontaneous magnetization to be anomalously suppressed
when $J_i\propto\exp(-L_2/\xi_2)$ is small, even well below $T_c\approx
T_{c1}$.
The suppression is easily estimated by combining the RG equation for the order
parameter
$dm/dl=-\omega'm$, where the scaling dimension \cite{Pokr_Patash}
\begin{equation}
\omega'=k_BT/4\pi\Gamma_1'=k_BT/4\pi\Gamma_1L_1,
\end{equation}
with the additive growth of $J_i$ given by (\ref{additive}); note that in the
case of a
superfluid the helicity modulus $\Gamma$ is proportional to the superfluid
density
$\rho_s$. The result is
\begin{equation}
\ln \frac{m(T)}{m_1(T)}=
-\frac{k_BT}{8\pi\Gamma_1(T)\xi_2(T)}\frac{L_2}{L_1}+O(L^{-1}).
\label{mXY3D}
\end{equation}
Note that $k_BT/\Gamma_1$ essentially plays the role of $\xi_1$ in this
universality
class, as well. One can thus see that in the case of $L_1/L_2=O(1)$,
spontaneous magnetization of a superlattice decreases exponentially as soon as
$\xi_1$ exceeds $\xi_2$, long before $\xi_1$ becomes comparable to the
thickness $L_1$. It would be interesting to see if this behavior could be
observed at a
superfluid transition of liquid helium-4 in some matrix with periodically
modulated
properties.

Similarly, a $d=2-XY$ superlattice may be realized in a helium
film deposited on a periodically modulated substrate
(the latter could be itself cut out of a {\em solid} three-dimensional
superlattice). In
this case the $d'=1$ $XY$-model orders only at $T=0$. Thus we expect $T_c$ of
the
superlattice to be shifted well below $T_{c1}$. At such temperatures
the 1-layers are equivalent to a
sequence of one-dimensional (classical) $XY$-chains, described by the Gaussian
Hamiltonians
\begin{equation}
{\cal H}_j=\int dx \frac{1}{2} \Gamma' (\nabla\phi_j)^2,
\label{TcXY2D}
\end{equation}
where $\phi$ is the phase of the $XY$ order parameter, $j$ labels the layers.
The effective one-dimensional helicity modulus $\Gamma'$ is given by the
product $\Gamma_1L_1$, as in all other cases considered in this section. Each
pair of
neighboring layers is coupled through the 2-layers separating them via
\begin{equation}
{\cal H}_{j,j+1}=J_i \int dx \cos (\phi_j-\phi_{j+1}).
\end{equation}
Combining the flow of the one-dimensional Gaussian coupling
$d\Gamma'/dl=-\Gamma'$ with the ubiquitous (\ref{additive}) ($d'=1$ here)
we arrive at the condition of criticality $\Gamma_1L_1J_i/(k_BT)^2=O(1)$,
which with logarithmic accuracy can be rewritten as
\begin{equation}
L_2/\xi_2(T_c)=\ln(L_1)+O(1).
\end{equation}
A little thinking shows that for $L_1/L_2=O(1)$ the critical temperature of the
superlattice is in fact shifted closer to the {\em lower} of the critical
temperatures of the two components; it stays however sufficiently above
$T_{c2}$ so that
$\xi_2$ remains much smaller than $L_2$ and our focusing on fluctuations in the
1-layers
remains well justified.

\section{Summary}

In summary, this paper addresses the following, as yet hypothetical,
experimental
project: A sequence of superlattices is built of the
same two ferromagnetic components, but with different basic thicknesses $L_1$
and $L_2$ (Fig. 1). The bulk critical properties of the components are assumed
to be
well known, the thicknesses $L_1, L_2\gg a$, i.e. large compared to any
important
microscopic length of the two materials, and no long-range dipolar
interaction is present. Given the well developed theory
of critical phenomena in the Ising, $XY$, and Heizenberg universality classes,
can one predict the variation of the critical properties of the superlattices
with
the changes in $L_1$ and $L_2$?

Naively one may expect that in the limit of $L_1,L_2\rightarrow\infty$ all
extensive
properties
of the superlattice can be expressed as a weighted average of the {\em bulk}
densities of the components. Indeed, if the limit of infinite thicknesses is
taken at
{\em fixed temperature} $T\neq T_c$, surface corrections to the bulk density
average vanish as $(\xi_1+\xi_2)/L$, where the correlation lengths
$\xi_1,\xi_2$ play
the role of {\em
penetration depths} for the perturbations induced by an interface in the two
adjacent layers. However, experiment on a given sample studies the
$T\rightarrow T_c$ limit at {\em fixed thicknesses} $L_1, L_2$, so that the
correlation length(s) are bound to exceed the period of the superlattice making
the surface contributions crucial. The thick layer limit for the critical
properties of a superlattice is {\em singular}, the thicknesses $L_1,L_2$ have
to be combined with certain functions of the temperature to form proper scaling
fields in terms of which a consistent scaling description is achieved.

Such scaling description has been constructed above based on
the qualitative picture of reentrant dimensional crossover \cite{onsager}.
In this picture (Fig. 2) a superlattice, at its critical temperature $T_c$,
displays three different asymptotic types of critical behavior,
when probed on different length scales. These three asymptotic models are,
in the order of increasing wave length, (ii) a sequence of noninteracting
semi-bulk samples, exhibiting {\em ordinary} surface critical behavior at the
surfaces; (iii) a sequence of noninteracting {\em thin} layers,
exhibiting $d'=d-1$-dimensional behavior; (iv) a highly anisotropic, but
uniform
critical system of the same universality class as that of the components.
Away from $T_c$ the finite correlation length falls into one of the three
regimes, so that instead of probing different length scales at $T_c$ one can
study the crossover by observing the asymptotic behavior at different
temperatures approaching $T_c$.

Each of these limiting types of behavior corresponds to a previously studied
renormalization group fixed point. Combining those into a global RG flow (Fig.
4)
we have constructed a rather complete scaling picture of critical phenomena in
superlattices composed of thick layers of two components.
The simplest case is that of a $d=3$-Ising superlattice, characterized by
the possibility of a finite-temperature phase transition in a single layer
of the first component. As a result, in the limit of thick layers the critical
temperature $T_c$ of the superlattice approaches that of the stronger of the
two ferromagnets, $T_{c1}$. The two crossovers then happen in a (relatively)
narrow vicinity of $T_c$. They are conveniently characterized in terms of the
reduced
temperature scales $\tau_D, \tau_R$ given by (\ref{tauD}), (\ref{tauR}).
These scales represent (Fig. 3) the relative widths of the temperature domains
in which the
superlattice behaves, respectively, as models (ii) and (iii) above. The
simplest scaling
description (\ref{A}) is then achieved in terms of the reduced temperatures
$t_D=(T-T_c)/\tau_DT_c$ and $t_R=(T-T_c)/\tau_R\tau_DT_c$,
shifted to the observed critical temperature $T_c$. As represented graphically
in the double logarithmic plot of Fig. 3, the ultimate divergence is
characterized by
the critical exponents characteristic of the bulk, $d$-dimensional behavior.
However
the presence of the intermediate domain of the $d'$-dimensional behavior (model
(ii))
leads to anomalous changes in the
critical amplitudes, as seen in (Fig. 3) and expressed analytically through
(\ref{tauD},\ref{tauR}) in (\ref{logCratio}).

The situation is somewhat more interesting in the other three- and
two-dimensional
spin-vector universality classes, where
the weak bonds between subsequent 1-layers play a crucial role in
maintaining the magnetic long-range order in the system. This leads to a large
shift
in the critical temperature of a superlattice, as implicitly given for
various classes by the universal expressions (\ref{Tc2D},\ref{TcH3D},
\ref{TcXY2D}),
as well as the anomalous, exponential suppression in the
amplitude of spontaneous magnetization (\ref{mXY3D}) in a three-dimensional
$XY$-superlattice. Where applicable, the results obtained here by the method of
this paper agree with the analysis of the exact solution for the
two-dimensional Ising superlattice \cite{Fish_Ferd} given in \cite{onsager}.
The three-peaked form of the temperature dependence of the specific heat,
characteristic of that model generalizes to the three-dimensional Heizenberg
class.
It has a transparent physical interpretation: most of the spin degrees of
freedom in such a superlattice get ordered at one of the bulk critical
temperatures of the two components, leading to finite, but large-amplitude
peaks at both. However, a small, $O(L^{-1})$, fraction of the degrees of
freedom participate in strong large-scale fluctuations at all
$T_{c2}\lesssim T \lesssim T_{c1}$, giving small-amplitude, but diverging
contribution to the specific heat at $T_c$.

The abundance of interfaces in a three dimensional magnetic
superlattice may greatly improve chances of experimental observation of surface
and two-dimensional scaling. In particular, the elusive surface scaling
exponents enter the definition of the parameter $\tau_R$ in (\ref{tauR}) and
can be obtained, for instance, by analyzing the dependence of the critical
amplitudes, (\ref{logCratio}) and Fig. 3, on the thickness of 1-layers $L_1$.

Finally, the main restriction of the present analysis seems to be the neglect
of dipolar interactions. When present, these interactions will lead to
demagnetization effects making Heizenberg universality class asymptotically
unstable. Otherwise, most of the results presented above can be easily modified
in the case when the interlayer interaction dependence on the thickness of the
separating layers, $J_i(L_2)$, is a power-law rather than an exponential as was
always assumed above. However, the assumption of local interactions enters our
consideration in much deeper ways than just through the exponential form
of $J_i(L_2)$, so that the required revision could be much more serious.
{\em Anti}-ferromagnets, of course,
are free from the complications arising from the dipolar interactions. However,
pinning of the spin density waves by the interfaces between the layers
in a superlattice gives rise to a different source of the thickness dependent
effects beyond the scope of this paper. In any case, the above consideration
seems to represent a reasonable first step towards a complete analysis of
scaling in superlattices. Note that recently critical phenomena in
superlattices
have been analyzed in the frames of the mean-field Landau-Ginzburg
approximation \cite{Hu Kawazoe}. A direct comparison of our results to theirs
is hardly
feasible in view of the failure of the mean-filed approximation $d=2$ and 3
considered here. However a combination of the results of
Ref. \cite{Hu Kawazoe} with a renormalization group expansion in $d=4-\epsilon$
and the scaling framework developed above looks promising, although
computationally very hard, direction for future work.

\acknowledgments
Comments by M. E. Fisher have been highly appreciated.

\appendix
\section{Interlayer coupling}

Here we derive the effective spin-spin coupling (\ref{int}) from the
Ornstein-Zernike
spin-density functional
\begin{equation}
F_b[{\bf s}({\bf r}]=\frac{k_BT}{2\chi_2}\int_{0}^{L_2}dz\ d^{d'} x'
\{{\bf s}^2+(\xi_2)^2(\nabla {\bf s})^2\}
\label{OZ}
\end{equation}
describing a layer of the second component on the length scales exceeding the
correlation length $\xi_2$; the other parameter entering (\ref{OZ}) is the bulk
spin
susceptibility $\chi_2$. This functional corresponds to the Gaussian RG fixed
point
describing the bulk paramagnetic phase of the second magnet. In a layer of
finite
thickness $L_2$, the bulk part of the functional has to be complemented by the
surface
contributions, which at the paramagnetic fixed point take the form
\begin{equation}
F_s=\sum_{j=1,2}F_{sj}=\\
\sum_{j=1,2}\int d^{d'}x'
\{\frac{1}{2}b\ {\bf s}^2({\bf x'},z_j)-{\bf h}_j({\bf x'}){\bf s}({\bf
x'},z_j)\}.
\end{equation}
In the above equation $z_1=0,\; z_2=L_2$, the phenomenological parameter $b$
accounts
for the change in the number of nearest neighbors for the surface spins, while
the
surface fields ${\bf h}_j$ are induced by the neighboring spins in the layers
of the
first component:
\begin{equation}
{bf h}_j({\bf x}')=J_{12}\ {\bf s}_j({\bf x}'),
\label{J12}
\end{equation}
where ${\bf s}_1={\bf s}_n^T$ and ${\bf s}_2={\bf s}_{n+1}^B$, as defined after
(\ref{int}), while $J_{12}$ is the effective coupling across the interface. The
minimization of the total free energy $F_b+F_{s1}+F_{s2}$ proceeds via Fourier
expansion in the $x'$-plane. For each Fourier harmonic of the spin density one
obtains an independent one-dimensional functional. Minimizing the latter and
summing all contributions one obtains in the
limit $\exp(-L_2/\xi_2)\ll 1$ an effective interaction between the surface
spins
of the first component
\begin{eqnarray}
-k_BT\sum_{\bf k'} G(k')\ {\bf h}_1({\bf k}'){\bf h}_2({\bf k}');\\
G(k')=\frac{(1+(k'\xi_2)^2)^{1/2}\xi_2/\chi_2}
{[b+(1+(k'\xi_2)^2)^{1/2}\xi_2/\chi_2]^2}.
\end{eqnarray}
Since the Fourier components of the spin density with wave numbers
$k'\gtrsim\xi_2^{-1}$ has been already eliminated by the RG transformation,
the factors $(1+(k'\xi_2)^2)^{1/2}\approx 1$. Using (\ref{J12}) one finally
obtains
the interaction of the form (\ref{int}) with the amplitude
\begin{equation}
J_0=\frac{2\xi_2/\chi_2}{[b+\xi_2/\chi_2]^2}J_{12}^2.
\end{equation}

\begin{figure}
FIG. 1: Schematic view of a superlattice in, a), $d=3$ and, b), $d=2$
spatial dimensions.
\end{figure}

\begin{figure}
FIG. 2: Graphical representation of the reentrant dimensional crossover. The
observed
behavior of the system changes from a noncritical superlattice, (i), to the
sequence of {\em thick} layers of the first compnent, (ii), connected by
exponentially
weak links mediated by noncritical fluctuations in the second component. From
this
limit, the direct crossover takes the system to a sequence of weakly coupled
{\em thin} layers, (iii), while the reentrant crossover takes it back to the
bulk,
although highly anisotropic, critical behavior, (iv).\end{figure}

\begin{figure}
FIG. 3: Schematic dependence of the specific heat per unit volume $C$ on the
scaled
temperature $t=(T-T_c)/T_c$ in a three-dimensional Ising superlattice.
\end{figure}
\begin{figure}
FIG. 4: Reentrant renormalization group flow in the {\em critical manifold} of
the
parameter space representing the $d=3$ Ising superlattices. The three fixed
points, defined in the text, are connected by critical separatrices (bold
lines).
They correspond, in the order in which they are encountered by the flow, to the
limits depicted by (ii), (iii), and (iv) in Fig. 2. The direct crossover,
from $CFP_s$ to $CFP'$, and the reentrant one, from $CFP'$ to $CFP_b$, are
driven by
two scaling fields, measured by the inverse thickness $L_1^{-1}$ and by the
interlayer
coupling strength $J_i$, respectively. Note that yet another relevant scaling
field,
$t$, directed roughly perpendicular to the plane of the figure,
drives the system away from this two-dimensional critical manifold to the high-
and low-tmeperature fixed points.
\end{figure}
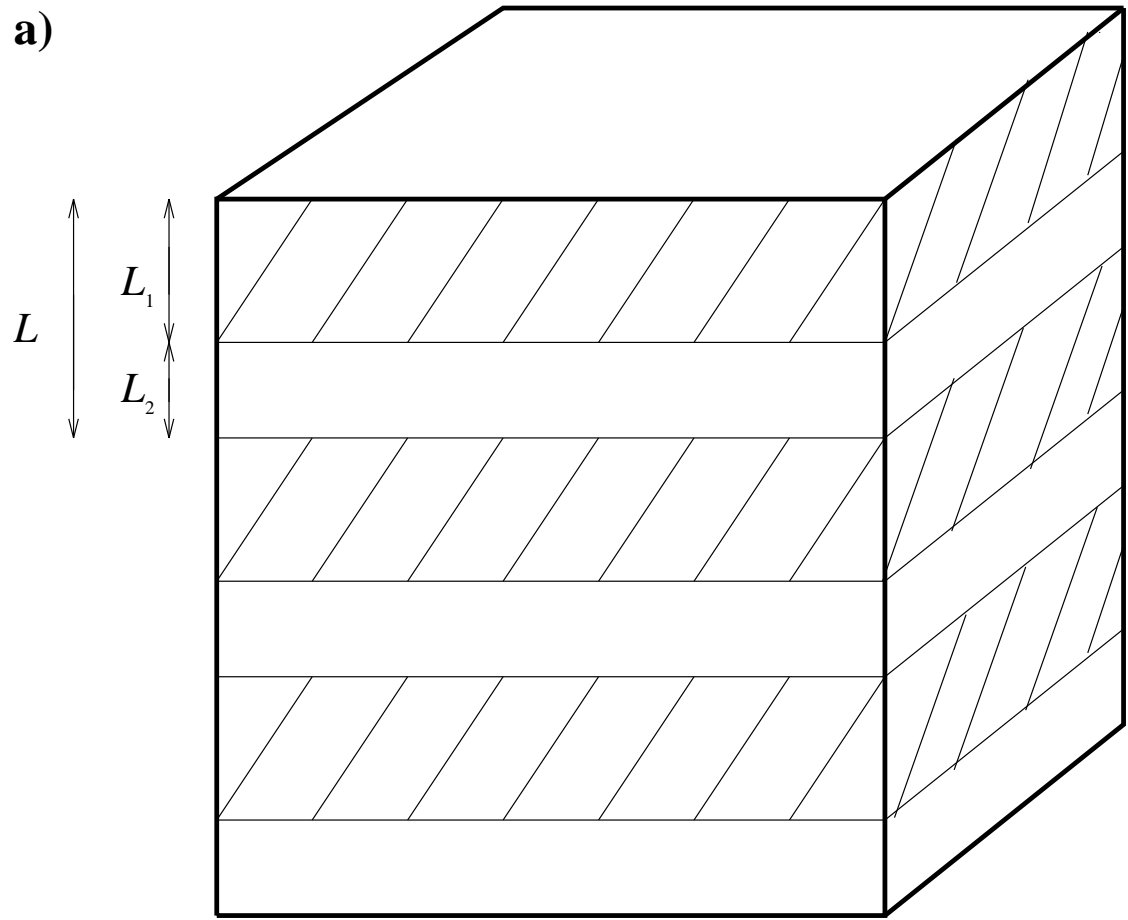
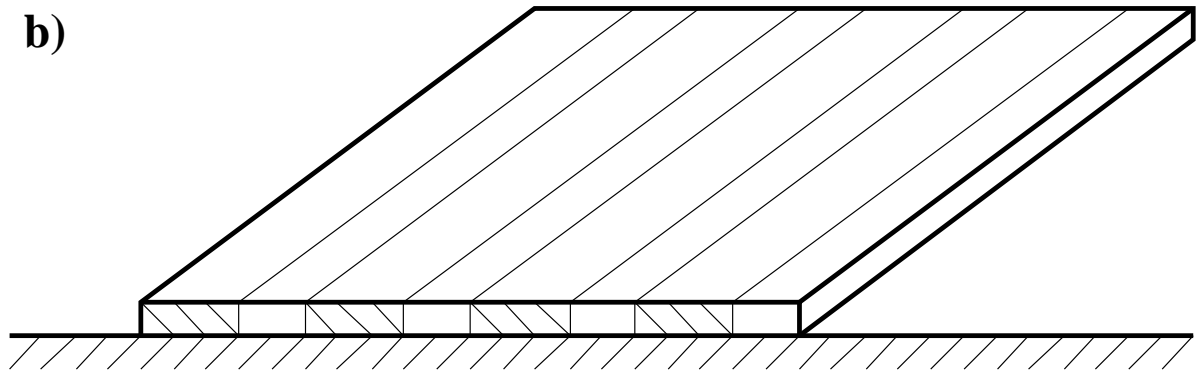
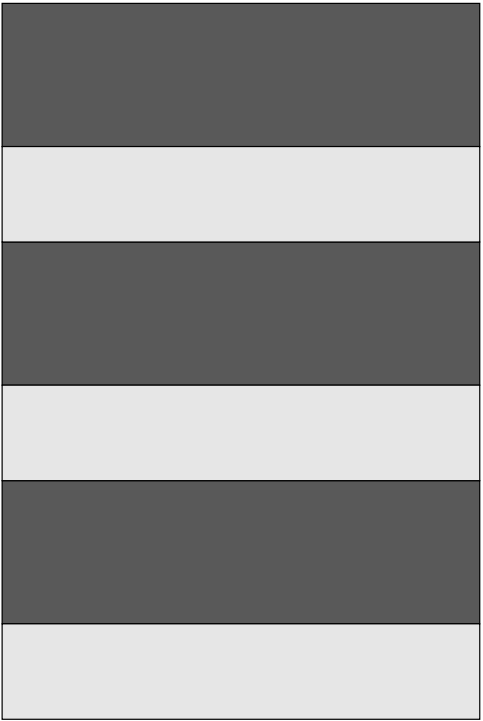
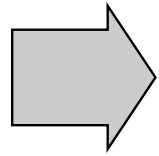
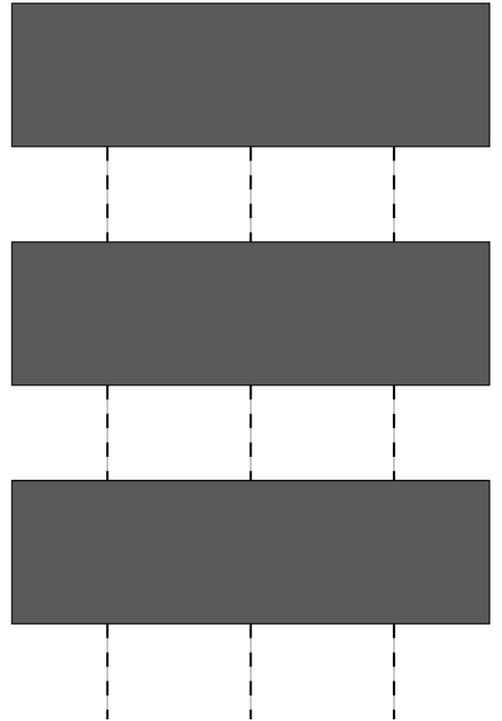
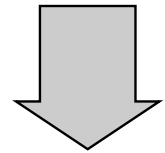
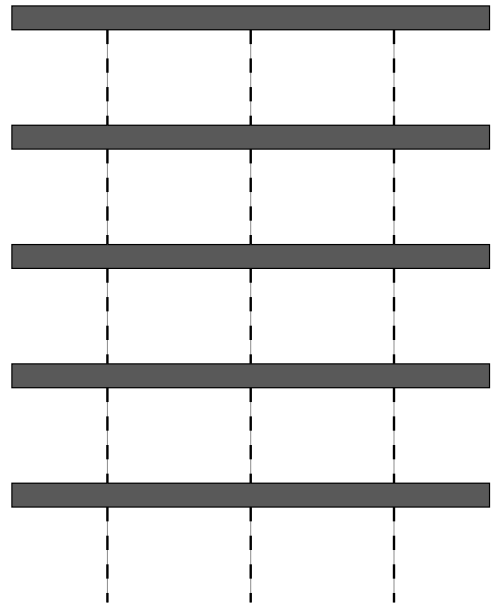
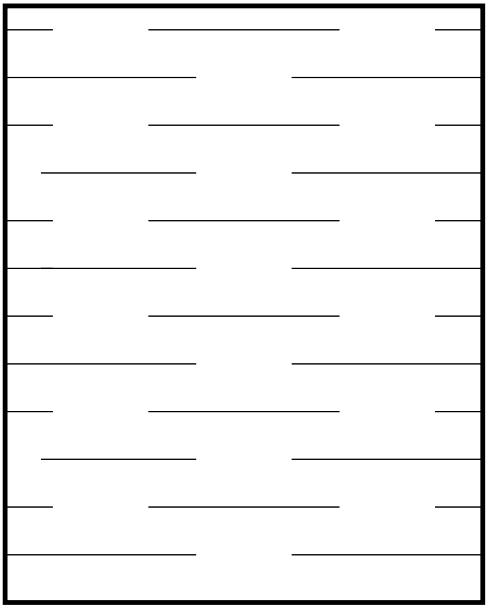
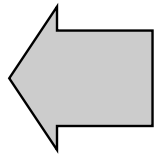
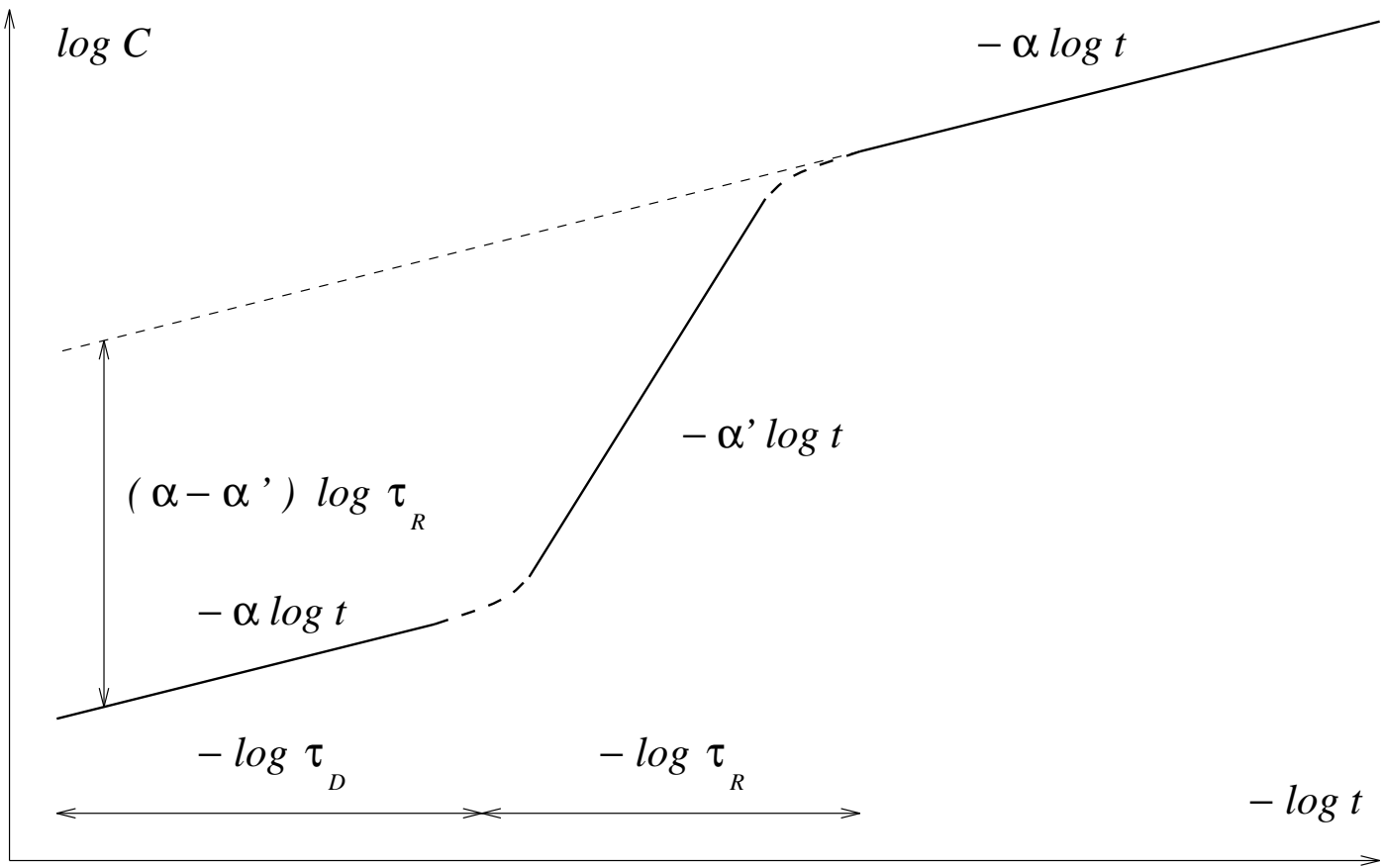
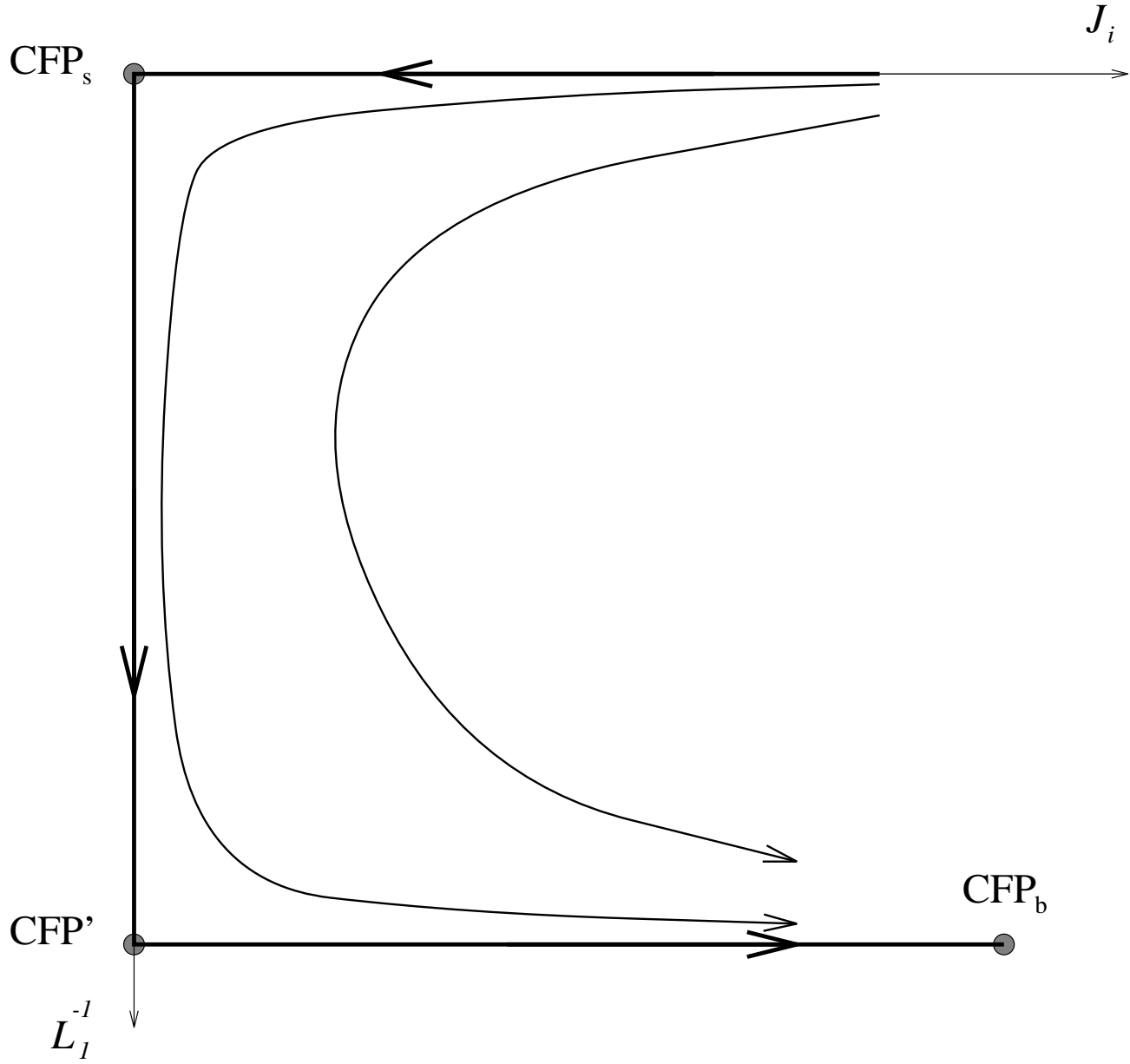
\end{document}